\documentclass[fleqn,twoside]{article}
\usepackage{espcrc2}

\usepackage{graphicx}
\usepackage[figuresright]{rotating}

\newcommand{\AmS}{{\protect\the\textfont2
  A\kern-.1667em\lower.5ex\hbox{M}\kern-.125emS}}

\title{Neutron Star Properties from an NJL Model Modified to Simulate
       Confinement}

\author{S.\ Lawley\address{Special Research Centre for the Subatomic
Structure of Matter, \\
University of Adelaide, Adelaide SA 5005, Australia} 
W.\ Bentz\address{Department of Physics, School of Science, 
Tokai University,
117 Kita-Kaname, Hiratsuka 259-1207, Japan}
        and
A.\ W.\ Thomas\address{Jefferson Lab, 12000 Jefferson Avenue, Newport News, 
VA 23606, U.S.A.}
       }
       
\begin{document}

\begin{abstract}
The NJL model has recently been extended with a method to simulate 
confinement. This leads in mean field approximation to a natural 
mechanism for the saturation of nuclear matter. We use the model to 
investigate the equation of state of asymmetric nuclear matter and then use 
it to compute the properties of neutron stars. 
\vspace{1pc}
\end{abstract}

% typeset front matter (including abstract)
\maketitle

\section{Introduction}\label{Intro}

The properties of isospin asymmetric matter at finite density are of 
particular interest in the study of nuclear physics and neutron stars.  
In particular, there is great interest in exploring whether stars with 
sufficiently high core density contain some form of 
quark matter~\cite{Glendenning:2000zz,Weber:2004kj,Panda:2003zj}.
In order
to investigate such questions it is natural to start with a model for 
dense matter which is built from the quark level. A number of models 
of this kind have been constructed, for example the quark-meson coupling 
(QMC) 
model~\cite{Guichon:1987jp,Guichon:1995ue,Saito:1994kg,Panda:2003ua,Chanfray:2002fk} 
and the chiral soliton model~\cite{Smith:2004dn}. 

We work 
within the framework of the flavor SU(2) Nambu--Jona-Lasinio (NJL) model, 
modified to simulate quark confinement.
As explained by Bentz and Thomas~\cite{BT},  
a description of nuclear matter can be obtained, which naturally exhibits 
the property of saturation, thus allowing for stable nuclei at low densities.  
The nucleon is constructed as a quark-diquark state by solving the Faddeev 
equation in the static approximation~\cite{Ishii:1993rt}.  
Using the proper time cut-off scheme to regularize the 
integrals~\cite{Hellstern:1997nv,Ebert:1996vx}, 
effectively introduces the phenomena of confinement, 
while leading to the stabilization of symmetric nuclear matter \cite{BT}.  
The general form for the equation of state is derived from the quark-level 
Lagrangian using the path integral formalism \cite{Bentz:2002um}.  
This model has been developed further to incorporate the phase transition 
to quark matter, including a color superconducting state~\cite{Bentz:2002um}.
However, in the present work we will deal with central densities 
where such transitions have not yet occured.

In this work we consider isospin asymmetric matter 
(neutrons, protons and electrons) in $\beta$-equilibrium.
Electrons are introduced to the system as a Fermi gas, 
balancing the positive charge of the protons.
We calculate neutron star masses, radii and profiles by solving the 
Tolman-Oppenheimer-Volkoff (TOV) equations \cite{TOV}.  
We find that this model can support stable stars up to a maximum 
mass of $2.19 M\odot$ where the central density is $0.97 fm^{-3}$.  
A typical mass neutron star ($1.4 M\odot$) has a central density 
of $0.4 fm^{-3}$, and a radius around 12.4 km.

\section{Model for the Equation Of State}\label{EOS}

The Lagrangian for the flavor SU(2) NJL model has the general form 
\begin{equation}
{\cal L} = \bar{\psi}(i\not\!\!{\partial}-m)\psi + \sum_{\alpha}G_{\alpha}(\bar{\psi}\Gamma_{\alpha}\psi)^2, 
\end{equation}
where $m$ is the current quark mass, $\psi$ is the flavor SU(2) 
quark field, $\Gamma_{\alpha}$ are the matrices in Dirac, flavor and 
color space, characterizing the chirally symmetric 4-fermi interaction 
between the quarks, and $G_{\alpha}$ are the coupling constants 
associated with these interactions. 
More explicitly, in matter of density $\rho$, we have 
\begin{eqnarray}
 {\cal L}&=& \bar{\psi}(i\not\!\!{\partial}-M-2G_{\omega}\gamma^{\mu}
\omega_{\mu}-G_{\rho}\gamma^{\mu}{\bf\tau\rm}.\rho_{\mu})\psi - 
\nonumber \\ && \frac{(M-m)^{2}}{4G_{\pi}} + G_{\omega}\omega^{\mu}
\omega_{\mu} + G_{\rho}\rho^{\mu}\rho_{\mu} +{\cal L}_I \, , \nonumber \\ 
\end{eqnarray}
where we define the effective quark mass (i.e., the constituent quark mass 
in-medium) as $M=m-2G_{\pi}\langle\rho|\bar{\psi}\psi|\rho\rangle$ and 
the vector and isovector fields in the nuclear medium are given by 
$\omega^{\mu}=\langle\rho|\bar{\psi}\gamma^{\mu}\psi|\rho\rangle$ 
and $\rho^{\mu}=\langle\rho|\bar{\psi}\gamma^{\mu}\tau\psi|\rho\rangle$.  
{}From this Lagrangian we derive both the equation of 
state and the quark-diquark structure of the bound ``nucleon''.
\begin{figure}[tb]
\begin{center}
\includegraphics[angle=-90,width=7.5cm]{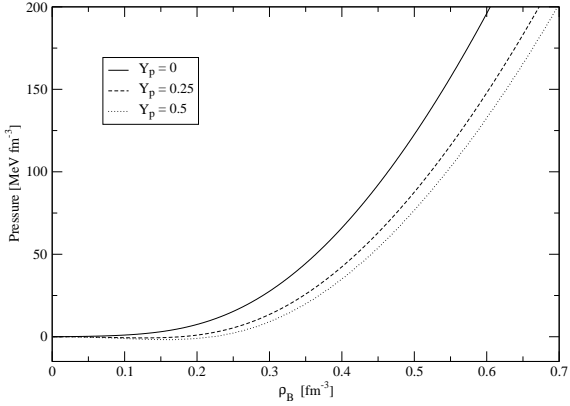}
\caption{Equation of state for asymmetric nuclear matter for various 
values of the proton fractions (Yp).}
\label{eos1}
\end{center}
\end{figure}
The interaction Lagrangian ${\cal L}_I$ consists of a sum of qq 
channel interaction terms, which are isolated using Fierz 
transformations~\cite{IBY}  
\begin{equation}
{\cal L}_I = {\cal L}_{I_{s}} + {\cal L}_{I_{a}} + \ldots
\end{equation}
In order to construct the nucleon state, we restrict ourselves to the 
scalar diquark channel (${\cal L}_{I_{s}}$) in the present calculation.
This will soon be extended to include vector di-quark correlations. 
The equation of state can be derived using the path 
integral formalism~\cite{Bentz:2002um}, or equivalent Lagrangian techniques~\cite{SW}.
The energy density is expressed as
\begin{equation}
{\cal E} = {\cal E}_{vac} +  {\cal E}_{N} + {\cal E}_{\omega} + 
{\cal E}_{\rho} ,
\end{equation}
where the vacuum term is given by
\begin{eqnarray}
{\cal E}_{vac} &=&  12i\int\frac{d^{4}k}{(2\pi)^{4}}ln
\frac{k^{2}-M^{2}+i\epsilon}{k^{2}-M_{0}^{2}+i\epsilon} + 
\frac{(M-m)^{2}}{4G_{\pi}} \nonumber \\
&-& \frac{(M_{0}-m)^{2}}{4G_{\pi}},
\label{E_vac}
\end{eqnarray}
\begin{figure}[tb]
\begin{center}
\includegraphics[angle=-90,width=7.5cm]{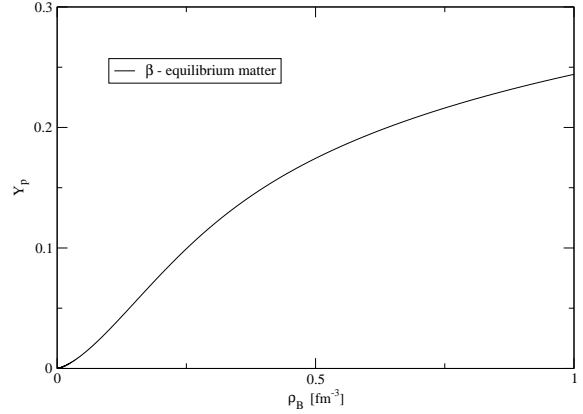}
\caption{Proton fraction for charge-neutral nuclear matter 
(neutrons, protons and electrons) in chemical equilibrium.}
\label{eos2}
\end{center}
\end{figure}
The Fermi motion of the nucleons contributes
\begin{equation}
{\cal E}_{N} = \sum_{N}\gamma_{N}\int\frac{d^{3}k}{(2\pi)^{3}}
\theta(k_{F_{N}}-k)\sqrt{k^{2}+M_{N}^{2}} \, ,
\label{E_N}
\end{equation}
where $M_N$ is the effective mass of the bound nucleon.
(This is calculated by solving the Faddeev equations using the 
in-medium masses of both the constituent quark and scalar di-quark.) 
The sum here is over neutrons and protons ($N=n,p$), 
which give equal contributions in the case of symmetric nuclear matter 
(of Fermi momentum $k_{F_{N}}$). 
\begin{figure}[tb]
\begin{center}
\includegraphics[angle=-90,width=8.5cm]{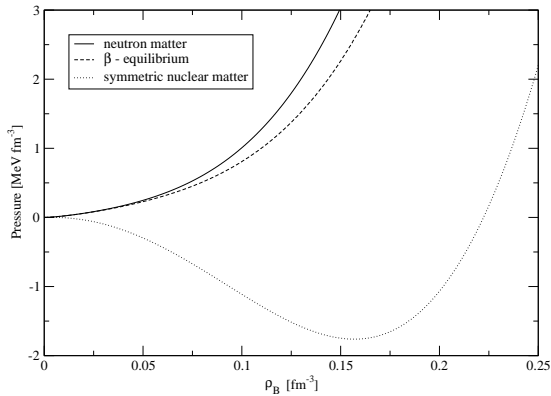}
\caption{Equations of state for neutron matter(solid line),
matter in ${\beta}$ - equilibrium (dashed line) and
symmetric nuclear matter (dotted line).}
\label{eos3}
\end{center}
\end{figure}
\begin{figure}[tb]
\begin{center}
\includegraphics[angle=-90,width=7.5cm]{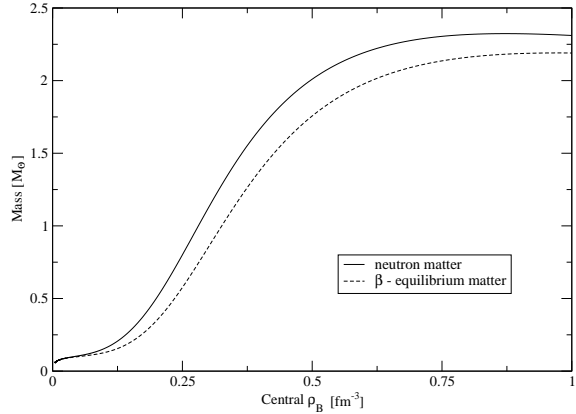}
\caption{Neutron star masses as a function of central baryon density.}
\label{star1}
\end{center}
\end{figure}
{}For nuclear matter at rest the omega meson contribution is given by, 
\begin{equation}
{\cal E}_{\omega}=9G_{\omega}\rho_{B}^{2},
\end{equation}
where the baryon density, $\rho_{B}$, is the sum of the proton and neutron 
densities,   
$\rho_{p}$ and $\rho_{n}$, respectively and $G_{\omega}$ is the 
vector coupling constant.  
The contribution to the energy density from the rho meson is given by,
\begin{equation}
{\cal E}_{\rho}=9G_{\rho}(\rho_{p}-\rho_{n})^{2} \, , 
\end{equation}
where the coupling constant $G_{\rho}$ is calculated at the empirical 
saturation density, $(\rho_{B},E_{B}/A)=(0.16fm^{-3},-15MeV)$, so that 
the symmetry energy coefficient is $32.5 MeV.$ 
\begin{figure}[tb]
\begin{center}
\includegraphics[angle=-90,width=8.5cm]{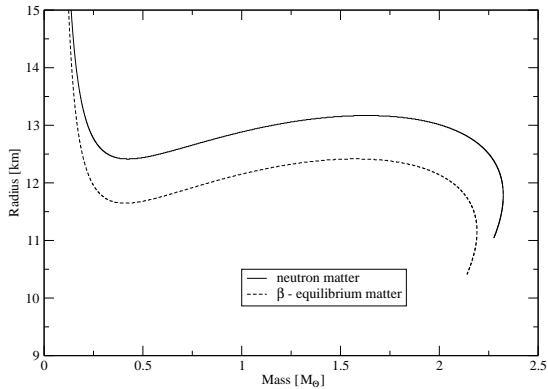}
\caption{Neutron star mass versus radius for pure neutron matter and for
nuclear matter in $\beta$-equilibrium.}
\label{star2}
\end{center}
\end{figure}
\begin{figure}[tb]
\begin{center}
\includegraphics[angle=-90,width=7.5cm]{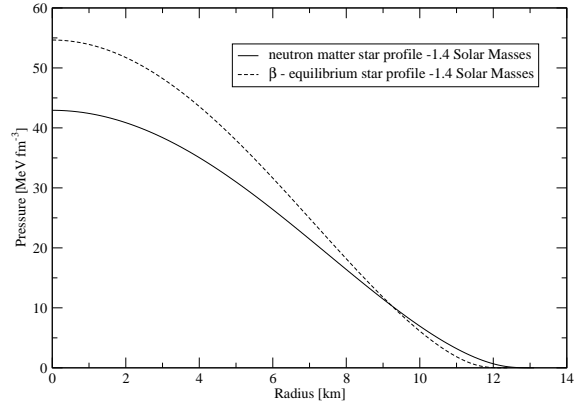}
\caption{Pressure profiles for neutron stars of mass
1.4 M${\odot}$.}
\label{star5}
\end{center}
\end{figure}
\begin{table}[b]
\caption{Parameters for the calculations of the properties of the 
nucleon and nuclear matter.}
\begin{center}
\begin{tabular}{l l}

\hline
\hspace{5mm}$G_{\pi} [GeV^{-2}]$ &\hspace{10mm}$19.60$ \hspace{5mm} \cr
\hspace{5mm}$G_{s}[GeV^{-2}]$ &\hspace{10mm}$0.51 G_{\pi}$\hspace{5mm} \cr
\hspace{5mm}$G_{\omega} [GeV^{-2}]$&\hspace{10mm}$0.37 G_{\pi}$\hspace{5mm}\cr
\hspace{5mm}$G_{\rho}[GeV^{-2}]$&\hspace{10mm}$0.092 G_{\pi}$\hspace{5mm}\cr
\hspace{5mm}$\Lambda_{UV} [MeV]$ &\hspace{10mm} $638.5$\hspace{5mm}\cr
\hspace{5mm}$\Lambda_{IR} [MeV]$ &\hspace{10mm} $200$\hspace{5mm}\cr 
\hspace{5mm}$m [MeV]$ &\hspace{10mm} $16.93$\hspace{5mm}\cr \hline
\end{tabular}

\end{center}
\end{table}

{}For a given density the constituent quark mass $M$ 
in Eq.~(\ref{E_vac}) is determined by the condition 
$\partial {\cal E}/\partial M=0$, where the quark mass at zero density 
is $M_{0}=400MeV$.  The nucleon mass $M_{N}$ in Eq.~(\ref{E_N}) 
is a function of $M$, determined by the pole in the Faddeev 
equation \cite{BT}.  The energy density and pressure are calculated 
implicitly as a function of baryon density.  
As shown in Fig.~\ref{eos1}, the presence of protons softens the 
equation of state.  This effect leads to more compact neutron stars.
To find the equation of state for matter in $\beta$-equilibrium, 
we introduce electrons, ${\cal E}_{e}=k_{F_{e}}^{4}/4\pi^{2}$, and 
impose the conditions of charge neutrality ($\rho_{p}=\rho_{e}$) and 
chemical equilibrium ($\mu_{e}=\mu_{n}-\mu_{p}$).  Fig.~\ref{eos2} 
indicates the proton fraction for matter in $\beta$-equilibrium 
as a function of baryon density.

\section{Results and Discussion}

Given the relationship between ${\cal E}$ and $P$ 
we can integrate the TOV equations,
\begin{equation}
\frac{dP}{dr} = -\frac{G({\cal E}(r)+P(r))(4{\pi}r^{3}P(r)+M(r))}{r(r-2GM(r))}
\label{tove}
\end{equation}
\begin{equation}
M= 4\pi \int_{0}^{R} {\cal E}(r)r^{2}dr
\end{equation}
The stars generated from matter in $\beta$-equilibrium are similar to 
those composed of pure neutron matter.  Fig.~\ref{star5} illustrates 
the profiles of typical mass stars for each equation of state.  
In the case of matter in $\beta$-equilibrium the central density is a 
little less than 3 times normal nuclear matter density.  
This corresponds to a proton fraction of about 15\% (see Fig.~\ref{eos2}), 
in the centre of the star.

\section{Concluding Remarks}

In conclusion, we have seen that matter in $\beta$-equilibrium produces 
more compact stars than pure neutron matter.  The abundance of protons 
depends on the value of $G_{\rho}$, 
which in our case is relatively low.
While the differences are quite small, the effect is clear.
We have found that this model predicts densities of about $0.4 fm^{-3}$ 
in the core of typical mass neutron stars.  Results for masses and 
radii are in basic agreement with astrophysical data.

This work was supported by the Australian Research Council and DOE
contract DE-AC05-84ER40150,
under which SURA operates Jefferson Lab.

\end{document}